\documentclass[aps,prl,preprint,showpacs]{revtex4-1}

\usepackage{hyperref}
\usepackage{graphicx}
\usepackage{amsfonts,amsmath,amssymb}
\usepackage[latin1]{inputenc}
\usepackage{color}
\usepackage{empheq}
\usepackage{float}
\usepackage{tabularx}
\usepackage[squaren,Gray]{SIunits}

\graphicspath{fig/}

\definecolor{forestgreen}{RGB}{34, 139, 34}

\newcommand{\ket}[1]{|\ \! #1 \ \! \rangle }

\begin{document}

\title{Absolute Single Ion Thermometry}
\author{Vincent Tugayé}

\author{Jean-Pierre Likforman}

\author{Samuel Guibal}

\author{ Luca Guidoni}
\email{luca.guidoni@univ-paris-diderot.fr}
\affiliation{Université Paris--Diderot, Sorbonne Paris Cité,\\
Laboratoire Matériaux et Phénoménes Quantiques, UMR 7162 CNRS, F-75205 Paris, France}


\begin{abstract} 
We describe and experimentally implement a single-ion local thermometry technique with absolute sensitivity
adaptable to all laser-cooled atomic ion species.
The technique is based on the velocity-dependent spectral shape of a quasi-dark resonance tailored in a $J\to J$ transition such that the two driving fields can  be derived from the same laser source leading to a negligible relative phase shift.
We validated the method and tested its performances in an experiment on a single $^{88}$Sr$^+$ ion cooled in a surface radio-frequency trap.
We first applied the technique to characterise the heating-rate of the surface trap.
We then measured the stationary temperature of the ion as a function of cooling laser detuning in the Doppler regime.
The results agree with theoretical calculations, with an absolute error smaller than 100~$\mu$K at 500~$\mu$K, in a temperature range between 0.5 and 3 mK and in the absence of adjustable parameters. 
This simple-to-implement and reliable method opens the way to fast absolute measurements of single-ion temperatures in future experiments dealing with heat transport in ion chains or thermodynamics at the single-ion level.
\end{abstract}
\maketitle
Laser cooled trapped ions offer the opportunity for a precise quantum control at the single particle level \cite{Wineland:2013} that triggered the development of ion-based platforms dedicated to quantum information processing \cite{Blatt:2008,Monroe:2014}.
Cold ion based systems have also been proposed and applied for testing thermodynamics in the quantum regime \cite{Rossnagel:2014, Rossnagel:2016} and  quantum heat transport in network chains \cite{Lin:2011,Bermudez:2013,Freitas:2016, Ramm:2014}.
The latter applications need the development of thermometry techniques that should characterise in a short time the velocity distribution of a single-ion that possibly interacts with other ions in its direct environment.
Since the first demonstrations of laser cooling of trapped ions \cite{Neuhauser:1978, Wineland:1978},  experimental thermometry tools have been developed together with theoretical models predicting stationary temperatures, for instance in the case of Doppler cooling \cite{Wineland:1979, Wineland:1987a}.
Among the thermometry techniques for trapped ions we can distinguish between three large families.\\
\indent In a first approach the spatial distribution of a single ion in a trap is measured by acquiring a time averaged image of its fluorescence.
The knowledge of the stiffnesses of the trap and the measurement of the spot size (corrected for diffraction) allow for the retrieval of the ion energy distribution.
This technique, introduced in early experiments \cite{Neuhauser:1980}, has been recently implemented to reach milliKelvin sensitivity \cite{Norton:2011, Knunz:2012} and to measure anomalous heating in a surface trap \cite{Boldin:2018}.
Spatial thermometry is only adapted to single ion experiments and the limited knowledge of point spread function of the imaging system affects both the accuracy and the precision of the method that reaches its best performances in shallow traps.\\
\indent A second family of thermometry techniques exploits the sensitivity to the ion motional state of the optically addressed narrow vibrational transitions in a trap (either quadrupole- or Raman-addressed).
In the Lamb-Dicke regime this technique has been introduced by Monroe and co-workers to characterise the ground-state cooling of a Be$^+$ ion in a trap \cite{Monroe:1995} and allow for the measurement of the average vibrational occupation number $\bar{n}$ in a given vibrational mode.
This technique has been used for the first studies of the heating rates associated to ion traps \cite{Turchette:2000}; it addresses relatively low temperature ranges ($\bar{n}<5$) but can be extended towards higher temperatures measuring the collapse of Rabi oscillations observed in the carrier or blue sideband transitions \cite{Meekhof:1996,Sikorsky:2017}.
A more sophisticated version of thermometry that also requires coherent manipulation of motional states can be assimilated to these techniques and extends the temperature range up to room temperature \cite{Johnson:2015}.
While extremely powerful and precise, these techniques are only able to characterise normal modes of oscillation one by one.
In the specific cases of multi-ion chains used for studying heat transport or in the experiments about quantum thermodynamics this limitation can be a serious drawback.\\
\indent The third approach to thermometry, that we address here, is based on the modifications of the photon scattering rate induced by the Doppler effect and has been used for instance in the first demonstrations of laser-cooling of ions  \cite{Neuhauser:1978, Wineland:1978}.
More recently the velocity-dependent photon scattering of a dipole-allowed transition involving two electronic levels has been exploited by recording and analysing the transient photon scattering rate during the Doppler cooling of an ion (Doppler re-cooling technique \cite{Epstein:2007, Wesenberg:2007}).
This technique has been widely used to measure heating rates of ion traps, it has also been extended to ions with richer level structures (e.g. Ca$^+$ and Sr$^+$), either using incoherent repumping schemes \cite{Allcock:2010} or including the multilevel structure in the analysis of transient photon scattering \cite{Meir:2017, Sikorsky:2017}.
The relatively large linewidth of the usual cooling transitions ($\simeq 2\pi\times 20$~MHz) implies that Doppler re-cooling thermometry can address the measurement of relatively high temperatures ($T>1$~K).
However multilevel ions addressed by at least two laser beams may also display the so called ``dark resonance'' phenomenon  \cite{Alzetta:1976, Arimondo:1976} that may originate spectral features with narrow linewidths \cite{Siemers:1992}. 
These dark resonances depend on motional state and may indeed be used for cooling the ions below Doppler limit \cite{Morigi:2000, Roos:2000, Lin:2013, Lechner:2016, Allcock:2016}, for evaluating the ion excess micromotion \cite{Lisowski:2005}, and for thermometry purposes \cite{Peters:2012, Rossnagel:2015}.
The main advantage of cooling and thermometry methods based on dark resonances is their ability to address (or sense) all vibrational modes without need of multiplexing.
In the case of thermometry this ability allows for a fast probing of the velocity distribution of a single ion that may be interacting with rich environments (Coulomb crystals or thermal baths) \cite{Rossnagel:2015}.\\
\indent In this paper we present a new implementation of dark resonance thermometry and demonstrate its ability to measure absolute temperatures (accuracy and precision in the sub-mK range) in the absence of calibrations with other methods.
\begin{figure}[]
\includegraphics[scale = 0.6]{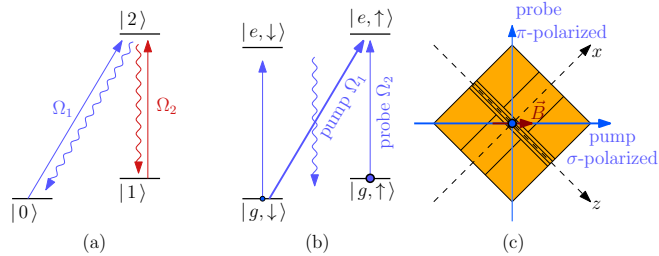} 
\caption{(a) energy level structure of a $\Lambda$-system that allows for the presence of a dark state; (b) quasi $\Lambda$-system: the presence of an additional level coupled to the light field reduces the lifetime of the dark state, but as long as $\Omega_2<\Omega_1$, most of the electronic population may accumulate in a quasi-dark state; (c) pump and probe beam geometry with respect to the trap axes used in the experiment to obtain a sensitivity to the temperature along the $z$-axis.}
\label{FIG_SETUP}
\end{figure}
This substantial improvement is obtained by addressing two transitions with a couple of laser beams that are only slightly detuned (some MHz) one with respect to the other.
This configuration eliminates phase stability issues that limited the precision in previous experiments \cite{Rossnagel:2015}.
The fit of the spectral experimental data with the solutions of the optical Bloch equations (OBE), in the absence of {\it ad hoc} parameters accounting for phase drifts of laser beams,
allowed us to obtain a measurement of absolute temperatures between 0.2 and 3~mK.
We also calculated that the range of applicability is adjustable up to 200~mK \cite{SupplInfo}.
In order to test experimentally this new thermometry tool we implemented it with a single Sr$^+$ ion loaded and cooled in a surface trap.
We exploited the measured temperatures to characterise the heating rate of the trap and to verify in an absolute way the theoretical law describing the stationary temperature reached with one-beam Doppler cooling \cite{Wineland:1987a}. 
We also demonstrated that this dark resonance approach is capable of cooling the ion well in the sub-Doppler regime (down to $T_z=0.13(1)$~mK) with a mechanism reminiscent of EIT cooling \cite{Morigi:2000} but in a completely different regime.\\
\indent Let us start considering a three level atom in a $\Lambda$ configuration driven by two laser fields with wave-vectors $\vec{k}_1$ and $\vec{k}_2$  (see figure \ref{FIG_SETUP} a)).
For an atom at rest, a fluorescence spectrum obtained scanning the detuning of one beam keeping the other detuning constant displays a dark resonance: a sharp drop in the scattering rate.
This feature is due to the presence of a ``dark state'':
\begin{equation}
\ket{\psi_D}(t) \propto \Omega_2\ket{0}-\Omega_1e^{i\varphi(t)}\ket{1}\\
\label{EQ_DARK_STATE}
\end{equation}
\noindent where $\varphi(t) = (\Delta_2-\Delta_1)t + \varphi_2(t)-\varphi_1(t)$ and $\Delta_i$, $\Omega_i$ and $\varphi_i(t)$ are the detunings, Rabi frequencies and the phases of the driving fields ($i=1,2$) \cite{Arimondo:1996, SupplInfo}.
As long as $\varphi$ does not depend on time $\ket{\psi_D}$ is not coupled to the driving fields: for $\Delta_1 = \Delta_2$ and if $(\varphi_2(t)-\varphi_1(t))$ is constant a dark resonance shows up because the electronic population is optically pumped in  $\ket{\psi_D}$.
For an atom that moves with velocity $\vec{v}$, the Doppler effect induces an additional  detuning $\delta_{i} = \vec{k}_{i}\cdot\vec{v}$  ($i=1,2$) giving:
\begin{equation}
\dot{\varphi}_{Doppler} = \delta_2-\delta_1 = (\vec{k}_1-\vec{k}_2).\vec{v}
\label{EQ_AXIS_SENSITIVITY}
\end{equation}
\noindent Atomic motion along the $\vec{k}_1-\vec{k}_2$ direction breaks the dark state condition and allows us to infer the velocity distribution from the contrast and shape of the dark resonance \cite{Peters:2012, Rossnagel:2015}.\\
\indent In a previous implementation of dark resonance thermometry with $^{40}$Ca$^+$ ions the frequency difference between the two driving lasers was very large (several hundreds of THz).
In this situation it may be very challenging to stabilise the relative phase of the driving fields: the contrast and shape of the dark resonance are affected by this technical issue that may be taken into account by an {\it ad hoc} parameter \cite{Rossnagel:2015}.
On the contrary, if the two lower levels of the $\Lambda$  system lie in the ground state manifold \cite{Peters:2012}, the two driving fields can be derived by the same laser source and their relative phase drift becomes negligible.
It is our approach: here we consider a generalised dark resonance in a quasi- $\Lambda$ system in which the two lower levels are Zeeman sub-levels of the ground state of an ion.
As shown in Fig.~\ref{FIG_SETUP}~b) a $J=1/2\to J^\prime=1/2$ transition is driven by a $\sigma^+$ polarised pump beam and a $\pi$-polarised probe beam with Rabi frequencies $\Omega_1$ and $\Omega_2$ respectively.
The quantification axis is imposed by a magnetic field $B$.
In this configuration, there is a residual coupling between the $\ket{g,\downarrow}$ state and the $\ket{e,\downarrow}$ that shortens the lifetime of $\ket{\psi_D}$ and thus reduces the contrast of the dark resonance.
However this is not a limiting factor because the choice of the values of $\Omega_1$ and $\Omega_2$ allows us to create a state arbitrarily close to a dark state.\\
\indent To obtain the theoretical lineshape of a fluorescence spectrum for an ion at rest in the configuration of Fig.~\ref{FIG_SETUP}~b), we calculate the ion density matrix  $\rho$ by solving the OBE in the stationary regime within the rotating wave approximation \cite{Cohen-Tannoudji:2011,SupplInfo}.
The scattering rate $S$ is given by $S = \Gamma_e\rho_{ee}$, where $\Gamma_e$ is the inverse of the lifetime of the excited state and $\rho_{ee}$ is the sum of the populations in the excited-state sublevels.
\begin{figure}[]
\includegraphics{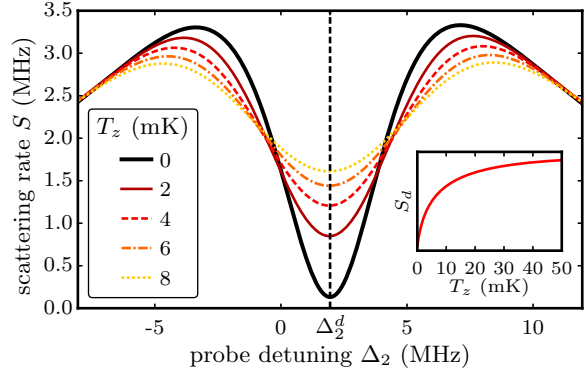} 
\caption{Calculated fluorescence spectra in the presence of a thermal velocity distribution for a motion along the $z$ axis and for several temperatures $T_z$.
The ion parameters correspond to the $5s\,\,^2S_{1/2}\to 5p\,\,^2P_{1/2}$ transition of the $^{88}$Sr$^+$.
The other parameters for the calculation are:
$\Omega_1 = \unit{12}{\mega\hertz}$, $\Omega_2 = \unit{4}{\mega\hertz}$, $\Delta_1 = \unit{-3.7}{\mega\hertz}$, $B = 2\times 10^{-4}$~Tesla.
Inset: scattering rate $S_d$ as a function of the temperature calculated with the previous set of parameters and $\Delta_2 =\Delta_2^d= +2$~MHz.
}
\label{FIG_Z_BROADENING}
\end{figure}
As shown in Fig.~\ref{FIG_Z_BROADENING}, (full thick line curve) a dark resonance occurs in the fluorescence spectrum ($S$ as a function of $\Delta_2$) for $\Delta_2=\Delta_2^d$ (two-photon resonance condition); the precise value of $\Delta_2^d$ depends mainly on the pump detuning $\Delta_1$, on the Zeeman shifts and, to a lesser degree, on the light shift induced by the probe beam.
We consider then an ion moving in an harmonic potential and, to simplify numerical calculations, we assume that the internal state evolution is much faster than the ion secular oscillation (weak binding approximation).
We describe classically the ion motional degrees of freedom and we consider a Maxwell-Boltzmann thermal velocity distribution described by a temperature $T_z$ such that $\hbar\omega_z \ll k_BT_z$; where $\omega_z$ is the secular frequency along the $z$ axis along which $\vec{k}_1-\vec{k}_2$ is aligned (Fig.~\ref{FIG_SETUP}~c)).
We also neglect the micromotion in $z$ direction (ideal linear trap). 
We then calculate the scattering rate for a moving ion as a two dimensional convolution of steady state solutions for an ion at rest with the distribution of velocity-dependent detunings $\delta_1$ and $\delta_2$.
Calculations show that, as expected, the motion along $\vec{k}_1+\vec{k}_2$ ($x$ direction) does not affect significantly the shape of the fluorescence spectrum.
On the contrary thermal motion along the $\vec{k}_1-\vec{k}_2$ ($z$) direction affects both the contrast and the width of the dark resonance as can be seen in Fig.~\ref{FIG_Z_BROADENING}.
In particular it is possible to calculate the scattering rate $S_d=S(\Delta_2=\Delta_2^d)$ as a function of the temperature $T_z$.
In this condition, as shown in the inset of Fig.~\ref{FIG_Z_BROADENING}, $S_d$ is a monotonous function of $T_z$. 
Therefore, the knowledge of the parameters $\Omega_1$, $\Omega_2$, $B$, $\Delta_1$ and $\Delta_2=\Delta_2^d$ allows us to calculate a temperature $T_z$ corresponding to a given scattering rate $S_d$.
In order to access different temperature ranges it is possible to tailor the width of the dark resonance by changing the pump and probe Rabi frequencies, while keeping constant their ratio (that fixes the contrast at zero temperature).
However, let us remark that narrower lines can only be obtained by lowering Rabi frequencies, then reducing the scattering rate.
Finally, a numerical analysis that relates the linewidth of the dark resonance, the scattering rate and the acquisition time necessary to reach a target statistical signal-to-noise ratio (SNR) affecting the temperature measurement shows that a range between  $2\times 10^{-2}$ and $2\times 10^{2}$~mK is accessible for a 20 s accumulation time and a SNR=10 \cite{SupplInfo}.\\
\indent The experiments are realised using single $^{88}$Sr$^+$ ions trapped in a symmetric five-wires surface trap \cite{Chiaverini:2005} with an ion-surface distance of 540~$\mu$m and $\omega_z\simeq 2\pi\times 100$~kHz.
\begin{figure}[]
\centering
\includegraphics{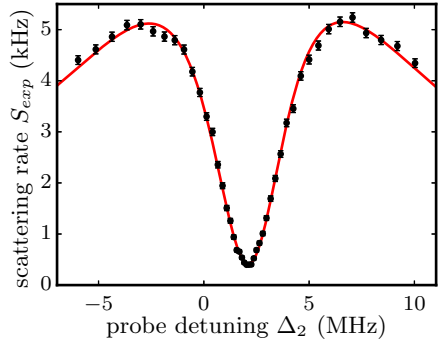} 
\caption{Averaged experimental fluorescence spectrum (black dots). The pump detuning is $\Delta_1 = \unit{-3,7}{\mega\hertz}$ and the ion is pre-cooled to a sub Doppler temperature with a cooling phase in which the probe is tuned to $\Delta_2 = \unit{1.7}{\mega\hertz}$. The free parameters obtained from the numerical fit (full red line) are $\Omega_2=3.78(1)$~MHz; $\Omega_1=10.7(1)$~MHz; $B=2.06(1)\times 10^{-4}$~Tesla; ion temperature $T=0.128(7)$~mK. The uncertainties do not take into account systematic shifts.}
\label{FIG_CALIBRATION}
\end{figure}
The details concerning the trap and the experimental setup can be found in ref.~\onlinecite{Szymanski:2012}.
We address the  $5s\,\,^2S_{1/2}\to 5p\,\,^2P_{1/2}$ transition; to avoid optical pumping in the $4d\,\,^2D_{3/2}$ we repump the electronic populations with an incoherent scheme \cite{Allcock:2010} that addresses simultaneously the $4d\,\,^2D_{3/2}\to 5p\,\,^2P_{3/2}$ and $4d\,\,^2D_{5/2}\to 5p\,\,^2P_{3/2}$ transitions.
The fluorescence spectra are acquired using a sequential approach to neglect the mechanical effect of the beams during the probing phase \cite{Gardner:2014, Meir:2014}.
We measure on a photon counter the scattering rate $S_{exp}=S \times \eta$ that is relied to the intrinsic scattering rate by the collection efficiency of the setup $\eta=1.95(5)\times 10^{-3}$.
\begin{figure}[]
\centering
\includegraphics{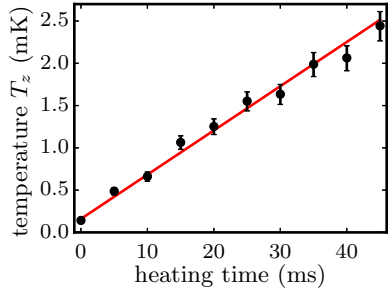} 
\caption{Averaged experimental retrieved temperature of the ion $T_z$ as a function of the heating time (black dots). After a preliminary phase of laser cooling the ion has a free evolution in the absence of laser beams during a variable time duration (heating time).  Let us remark that we imposed an initial sub-Doppler temperature ($T_{in}=0.20(5)$~mK) to extend the range of this measurement. The measured heating rate of the Trap (slope of the linear fit, full red line) is 52(1) mK/s. Error bars are $\pm$ one standard deviation (both statistical and systematic).}
\label{FIG_HEATING_RATE}
\end{figure}
The detunings $\Delta_1$ and $\Delta_2$ are imposed by two acousto-optic modulators that shift with negligible phase noise a common laser beam locked to an atomic reference with a precision better than 100~kHz.
As a first check we acquired several fluorescence spectra for different $\Omega_1$, $\Omega_2$ and $B$  for an ion cooled close to the Doppler limit.
Numerical fits with the solution of the OBE are in excellent agreement with the experimental spectra and the parameters extracted from the fits follow the variations imposed to the experimental parameters and give a constant temperature $T_z$ consistent with the Doppler limit.
Let us note that the sequential character of the acquisition is compulsory to sample the velocity-dependent scattering rate $S_{exp}$ and therefore, after averaging on several realisations, to retrieve what we call the temperature $T_z$ of the ion. 
For a fixed set of known parameters $\Omega_1$, $\Omega_2$, $B$, $\Delta_1$ and $\Delta_2=\Delta_2^d$, $T_z$ can be retrieved from a measurement of the average scattering rate $S_d$ (see above).
In the following we will adopt this approach, faster than the acquisition of a full fluorescence spectrum for each temperature.
In order to precisely evaluate the $\Omega_1$, $\Omega_2$, and $B$ parameters we acquire and fit a preliminary calibration spectrum, such as the one presented in Fig.~\ref{FIG_CALIBRATION}.\\
\indent As a first application we have measured the heating rate of the surface trap affecting the motion along the $z$ axis.
For that purpose we prepared the ion at an initial temperature $T_{in}$ and then measured its temperature after a heating phase during which laser cooling is switched-off.
The results of this measurements are shown in Fig.~\ref{FIG_HEATING_RATE} and display as expected a linear behaviour that allow us to measure an heating rate of $52(1)$~mK/s.
In order to extend the range of these measurements we pre-cooled the ion to a sub-Doppler temperature $T_{in}=0.20(5)$~mK (the Doppler limit for Sr$^+$ with the cooling scheme of Fig.~\ref{FIG_SETUP}  is $T_D=0.47$~mK) taking advantage of the mechanism reminiscent of EIT cooling that operates with $\Delta_2$ tuned on the low frequency side of the dark resonance.
This new cooling method occurs in a completely different regime with respect to previous realisations of EIT cooling \cite{Morigi:2000}; the complete description of the mechanisms involved is beyond the scope of this paper.
\begin{figure}[]
\centering
\includegraphics{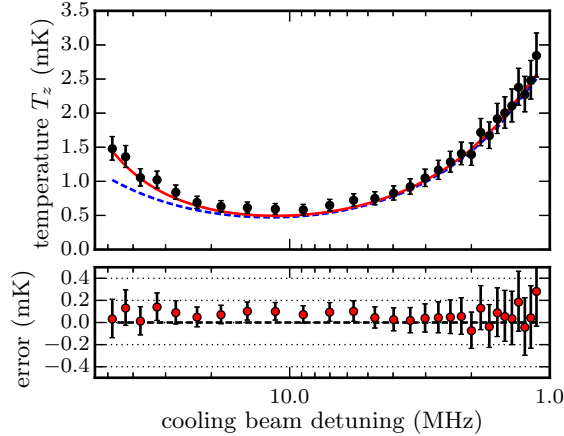} 
\caption{Averaged experimental retrieved temperature of the ion $T_z$ as a function of the Doppler-cooling beam detuning (black dots). Measurements are obtained in the low intensity regime  $I\ll I_{sat}$. The full red line (blue dashed line) is the result of calculation of the stationary temperature including (not including) the trap heating rate (see Fig.~\ref{FIG_HEATING_RATE}).
Let us remark that there are no free parameters in the calculations because both the scattering rate as a function of cooling beam detuning and the heating rate are the results of independent experimental measurements. The bottom of the figure shows the residuals (red dots) with the associated statistical standard deviation.}
\label{FIG_DOPPLER_LIMIT}
\end{figure}
To validate the absolute character of this thermometry technique we have measured the ion temperature as a function of the detuning of a Doppler cooling beam at low intensity \cite{Knunz:2012}.
Indeed in this regime it is possible to calculate analytical expressions describing both the cooling dynamics and the stationary temperature \cite{Wineland:1987a}.
The experimental sequence includes two cooling phases.
The first step has a duration of 20~ms and is performed at optimal detuning, allowing us to reach a temperature on the order of 0.5~mK with a theoretical characteristic time of 320~$\mu$s.
The second cooling step, performed at intensity $I\ll I_{sat}$ (where $I_{sat}$ is the saturation intensity), has a duration set to ten times the characteristic cooling time as calculated from the theory.
This characteristic time depends on the cooling beam detuning: the experimental sequences take into account this effect.
The measured temperature of the ion as a function of the cooling beam detuning is shown in Fig.~\ref{FIG_DOPPLER_LIMIT} (black dots).
In order to compare this result with the theoretical prediction, we acquire an experimental fluorescence spectrum (scattering rate as a function of cooling beam detuning) at the same (low) cooling beam intensity.
The sequential character of the acquisition and the incoherent repumping scheme allows us to describe the spectrum in terms of a pure Lorentzian line-shape.
We then inject the two parameters describing the Lorentzian scattering rate in the theoretical model of Doppler cooling and obtain the stationary temperature as a function of cooling beam detuning (dashed line in Fig.~\ref{FIG_DOPPLER_LIMIT}), in the absence of adjustable parameters.
The discrepancy with the experimental data observed at large detunings is dominated by the effect of the trap heating rate (see above), as shown by the stationary temperature obtained with calculations that include this term (full line in Fig.~\ref{FIG_DOPPLER_LIMIT}).
The error bars in Fig.~\ref{FIG_DOPPLER_LIMIT} ($\pm 1$ standard deviation) are calculated taking into account both photon counts statistics and calibration uncertainties \cite{SupplInfo}.\\
\indent In conclusion we demonstrated a new thermometry technique that we implemented with a single Sr$^+$ ion loaded and cooled in a surface trap.
We applied the technique to characterise the heating rate of the trap and to verify, in an absolute way, the theoretical law describing the stationary  temperature reached with one-beam Doppler cooling. 
The excellent agreement between the calculations and the experimental measurements over a large range of detunings demonstrates the reliability of the method and opens the way to fast (the probing time in a sequence is on the order of 10~$\mu$s) and directional single ion absolute thermometry.\\
\indent We thank M. Apfel, P. Lepert and M. Nicolas for technical support.
This study was partly founded by Région Ile-de-France through the DIM Nano-k (DEQULOT grant).

\end{document}